\documentclass[twocolumn,floats,showpacs,prl,amsmath,amssymb,floatfix,nofootinbib,balancelastpage]{revtex4}

\usepackage{epsfig}
\usepackage{color}

\begin{document}

\title{Detecting a physical difference between the CDM halos in simulation and in nature}

\author{Weike Xiao$^{\dagger}$, Chang Peng$^{\dagger}$, Xianfeng Ye$^{\dagger}$, Heng
Hao$^{\star}$ }

\affiliation {$\dagger$ Center for Astrophysics, University of
Science
and Technology of China,  Hefei, Anhui, 230026,P.R.China \\
$\star$ Astronomy Department, Harvard University, 60 Garden
Street, Cambridge, MA 02138}

\email{wkxiao@mail.ustc.edu.cn}
\date{\today}

\begin{abstract}

Numerical simulation is an important tool to help us understand
the process of structure formation in the universe. However many
simulation results of cold dark matter (CDM) halos on small scale
are inconsistent with observations: the central density profile is
too cuspy and there are too many substructures. Here we point out
that these two problems may be connected with a hitherto
unrecognized bias in the simulation halos. Although CDM halos in
nature and in simulation are both virialized systems of
collisionless CDM particles, gravitational encounter cannot be
neglected in the simulation halos because they contain much less
particles. We demonstrate this by two numerical experiments,
showing that there is a difference on the microcosmic scale
between the natural and simulation halos. The simulation halo is
more akin to globular clusters where gravitational encounter is
known to lead to such drastic phenomena as core collapse. And such
artificial core collapse process appears to link the two problems
together in the bottom-up scenario of structure formation in the
$\Lambda$CDM universe. The discovery of this bias also has
implications on the applicability of the Jeans¡¯ Theorem in
Galactic Dynamics.

\end{abstract}

\maketitle


\section{Introduction.}
\label{sect:intro}

Thanks to rapid development in numerical techniques, the N-body
simulation method has scored considerable success in furthering
our understanding of the formation and evolution of the
large-scale structures in the universe. But the simulations based
on the CDM models has not worked so well on small scales, one
aspect is the "cusp problem": The central density profile of the
simulation halos (Fukushige \& Makino 1997; Moore et al.1999;
Ghigna et al. 2000; Navarro et al. 1996) are too cuspy, and are
inconsistent with the observations on dwarf galaxies and low
surface brightness galaxies (Gentile et al. 2004; Primack 2003).
It is also inconsistent with the distribution of dark matter in
galaxy clusters found from X-ray and gravitational lensing studies
(Katayama et al. 2004; Tyson et al. 1998 ). Another aspect is that
$10\sim100$ times more substructures are predicted than is
observed (Willman et al. 2002; Hilker et al.2005);--- the
so-called "substructure problem".  Some alternative models were
introduced to deal with these problems, including the
self-interaction dark matter(SIDM) model, the warm dark matter
model etc.\ (Spergel \& Steinhardt 2000; G\"{o}tz et al. 2003;
Boehm et al. 2002).

Here we realize that our understanding of these two problems can
be greatly helped by studying a common, basic technical
difficulty: the huge difference between the number of particles
used in simulations of CDM halo, ($N_{halo}$) and the number
occurring in nature: even in the most powerful "high-resolution"
simulations today, a large dark matter halo contains only about
$10^6$-$10^7$ particles. Since we have to keep to a fixed mean
density of the universe, the mass of each particle in the
simulation has to be some 70 powers of ten the GeV candidates in
particle physics (Bertone et al.\ 2001; Bertone et al.\ 2005)

It is hard to accept that the mass of the CDM particle can be as
high as the mass of a small galaxy.  We have unavoidably
introduced a physical bias in our simulations which would impact
on small scale problems. The bias is that we have simultaneously
reduced the particle number density $n$ and raised the mass of
each particle $m$ by a huge factor, $f_{AE}\sim 10^{70}$. CDM
particle systems with such a physical bias can be expected to have
very much biased dynamical and statistical properties, and the
clustering and evolution scenario of CDM halos we learn from them
might be different from reality. This problem should be discussed
in more detail.


\begin{figure*}

\centerline{\epsfig{file=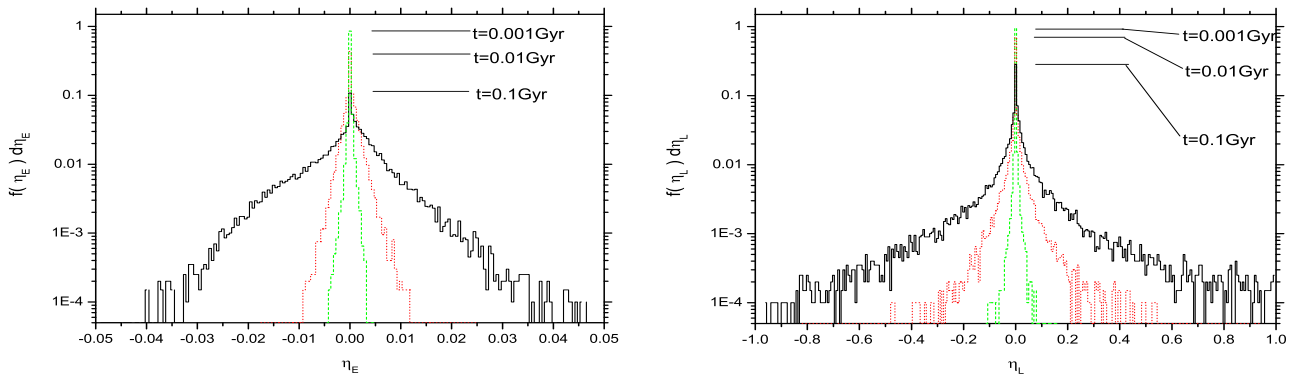,width=8in,height=3in,angle=0}}
\caption{ Time evolution of the distribution function
$f(\eta_{Ei},t)$ and $f(\eta_{Li},t)$ of a stable and spherical
halo. Different from the expected delta function of a halo in
nature, the $f(\eta,t)$ of the simulation halo in 0.001
Gyr(dash/green line), 0.01 Gyr(dot/red line) and 0.1
Gyr(solid/black line) is becoming more and more dispersive.}

\end{figure*}

\section{Relaxation And Detecting Method}
\label{sect:relaxation}

How will such bias affect the simulation results ?  One effect is
it changes the mean free path of two-body relaxation, $L_s$. For a
virialized CDM halo with mass $M$ and radius $R$, one can roughly
estimate $L_s$ as (Xiao et al. 2004):

\begin{equation}
L_s \simeq N_{halo}\frac {R}{3} \frac{\overline{\rho}}{\rho}
\end{equation}
where $\overline{\rho} =M/(4 \pi R^3/3)$ is the mean density of
the halo, and $\rho$ is its local density. For a virialized halo
in nature that is composed of a huge number of $GeV$ particles,
it's easy to see from eq.(1) that $L_s\gg R$ and that we can
completely disregard the two-body relaxation. However, for a
virialized halo in simulation, composed of a much smaller number
of huge mass particles, we have $L_s \sim R$, and the
corresponding two-body scattering cross section is close to the
value expected in the SIDM model. So, in the simulation, the
effect of relaxation can no longer be neglected. Considering that
star clusters that suffer similar relaxation can have their
dynamical and statistical properties seriously affected, leading
to core collapse, evaporation, etc.\ (Binney \& Tremaine,1987),
the discovery of this physical difference between CDM halos in
simulation and in nature can help us understand the two small
scale problems encountered in our current simulations.

In an effort to quantify and detect the relaxation process we have
designed two numerical experiments for a spherical CDM halo.
Different from the traditional statistical method (Binney \& Knebe
2002; Diemand et al. 2004), we study the relaxation process from
the viewpoint of the individual particle, and we find that the two
integrals of motion, the energy and angular momentum of the $i$th
particle of the halo, $E_i(\mathbf{x},\mathbf{v})$ and
$L_i(\mathbf{x},\mathbf{v})=|\mathbf{L}|$, can be effective
detectors of the relaxation effect caused by the physical bias.

Although, a globular cluster and a stable spherical CDM halo in
nature are both composed of collisionless gravitational
interacting particles, their values of $L_s/R \sim N$ show that
their gravitational potentials on small scale are very different.
For a CDM halo in nature, gravitational encounter can be neglected
and the halo potential $\phi(r)$ is stable and spherical for each
CDM particle. Then $E_i$ and $L_i$ of each particle, being both
integrals of motion, will be constant in such a system. In
contrast, for a globular cluster or a CDM halo in simulation, $L_s
\sim R$, and this means the collisionless particles in these
systems have to experience gravitational encounters, and the
potential $\phi$ is neither spherical nor stable. Then the $E_i$
and $L_i$ of each particle are no longer integrals of motion and
are soon modified by encounters.--- (The energy $E$ and angular
momentum $L$ of the whole system are still the same. This is an
important requirement for the reliability of many simulation
results, but it is not relevant to the problem on hand.) From this
point of view, the CDM halo in simulation is more like a globular
cluster than a CDM halo in nature.

As measures of the relaxation effect in simulation CDM halos, we
now define two parameters,
\begin{equation}
\eta_{Ei}=\frac{\Delta E_i}{|E_{i0}|}  \;\;and \;\;
\eta_{L_i}=\frac{\Delta L_i}{|L_{i0}|}
\end{equation}
where $\Delta E_i=E_i(t)-E_{i0}$ and $\Delta L_i=L_i(t)-L_{i0}$
are the variations of $E_i$ and $L_i$ from their initial values
$E_{i0}$ and $L_{i0}$. For a CDM halo in nature, the distribution
functions will simply be the time-independent delta-functions; for
a stable, spherical halo in simulation, in contrast, they will be
very different and will evolve in time.

\section{Simulation And Result of Numerical Experiments}
\label{sect:simulation}
Here we present two controlled numerical experiments using the
publicly available N-body tree code, GADGET (Springel et al.,
2001), to detect the relaxation effect discussed above. Following
Ma et al.\ (Gozdziewski et al.2005; Ma \& Michael 2004; Michael \&
Ma 2004), we generate a stable spherical virialized CDM halo with
a given NFW density profile (Navarro et al.1996):
\begin{equation}
\rho(r)=\rho_{crit}\overline{\delta}/[x(1+x)^2]
\end{equation}
where $x=r/r_s$,$\overline{\delta}=200c^3/3[ln(1+c)-c/(1+c)]$, and
$c=r_{vir}/r_s$ is the ratio of the halo's virial radius to scale
radius. In all the runs, we set $r_s=16kpc/h$, the total mass of
the halo $M=10^{12}M_\odot$, and we use a force softening of
$\epsilon=2kpc/h$.
\footnote{Our test runs showed that different values of $\rho(r)$
and $\epsilon$ etc.\ do have a small effect on the results, but
will not change the general qualitative picture.}
We drew the particle velocities from a local isotropic Maxwellian
distribution with the radius-dependent velocity dispersion
computed from the Jeans equation.

\begin{figure}

\centerline{\epsfig{file=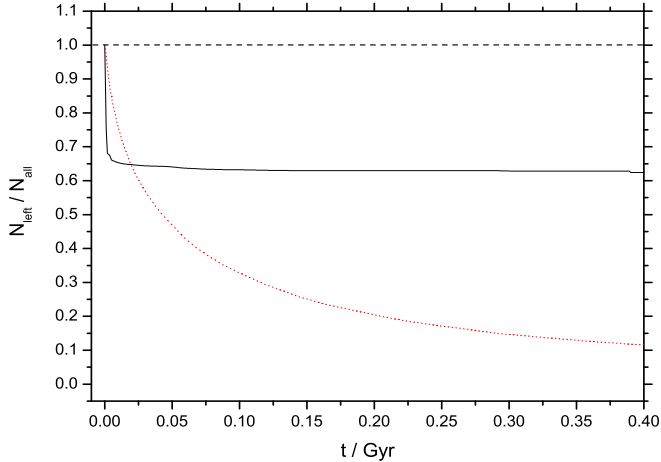,width=4in,height=3in,angle=0}}
\caption{ Time evolution of $N_{left}/N_{all}$ for $E_i$
(solid/black line) and $L_i$ (dot/red line). We can see the $E_i$
and $L_i$ have been changed for most of the particles in a short
time, different from the expected dash line of the halo in nature
that will not be affected by such relaxation.}
\end{figure}

In the first experiment, we start with a stable halo of
$2\times10^4$ particles. We first let the individual particles of
the NFW halo evolve for 5.0 Gyr, and we use the evolved halo as
our initial condition, noting down the values $E_{i0}$ and
$L_{i0}$ of each particle at this time. Then we record the
evolution of $E_i$ and $L_i$ of each particle and obtain the
distribution functions $f(\eta_{Ei}, t)$ and $f(\eta_{Li}, t)$ at
chosen epochs, $t$. Theoretically, as integrals of motion, the
$E_i$ and $L_i$ of each particle of such a halo in nature will be
constant and the distribution function $f(\eta)$ should remain a
delta function all the time. However, as we can see in Fig 1, both
$f(\eta_{L_i})$ and $f(\eta_{E_i})$ are spreading out as time goes
on. The spreading means that within a period of 0.1 Gyr, more and
more particles have experienced changes in their $E_i$ and $L_i$.

We define $N_{left}$ as the number of particles which satisfy
$|\eta_{Ei}|<0.01$ or $|\eta_{Li}|<0.01$ as measures of the
evolution of the distribution functions. As we can see in Fig.2,
for most particles in the simulated halo, $E_i$ and  $L_i$ were
changed very soon, and that the changes were faster in $L_i$ than
in $E_i$. In contrast, the value of $N_{left} /N_{all}$ remains
constant at 1.0 for a CDM halo in nature that consist of a huge
number of GeV particles.

In the second experiment, we consider a spherically evolving halo.
We multiply the velocity of each particle in the halo of the first
experiment by a factor of 0.9 as the initial velocity. Then the
halo will still be spherically symmetric, but it is no longer
stable. In this case, $E_i$ will be changed by the violent
relaxation. Meanwhile, for a CDM halo in nature including numerous
GeV collisionless particles, the $L_i$ of each particle will
remain as an integral of motion and will not change. Fig 3 shows
the distribution function $f(\eta_{Li})$ of the simulation halo.
Similar to Fig 1, the $L_i$ of each particle changed rapidly due
to this hitherto unsuspected process of galactic relaxation.

\begin{figure}
\centerline{\epsfig{file=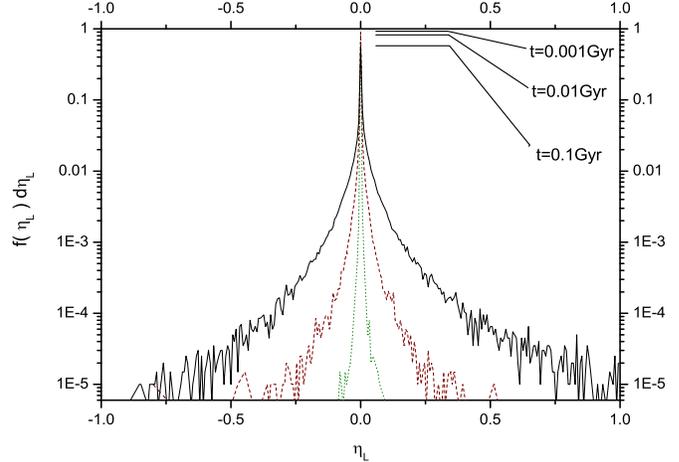,width=4in,height=3in,angle=0}}
\caption{Time evolution of the distribution function
$f(\eta_{Li},t)$ of a spherical evolving halo. Similar with the
stable halo in Fig 1, the $f(\eta,t)$ of the simulation halo in
0.001 Gyr(dash/green line), 0.01 Gyr(dot/red line) and 0.1
Gyr(solid/black line) is becoming more and more dispersive.}
\end{figure}

Finally, we tried to assess the effect of particle number on the
distribution functions. We generated three systems with,
respectively, $2\times10^4$, $2\times10^5$ and $2\times10^6$
particles, the other parameters remaining the same as for the
stable halo of the first experiment. Fig 4 shows the time
evolution of $\sigma_{\eta_L}^2\equiv\langle\eta_{Li}^2\rangle$.
We find that in all cases $\sigma_{\eta_L}^2$ increases as a
power-law function of time (within about $0.1 Gyr$), and, at any
given time, $\sigma_{\eta_L}^2$ is a fast decreasing function of
$N$. In other words, the relaxation effect, or the bias, is
greatly reduced when $N$ is increased.

\begin{figure}

\centerline{\epsfig{file=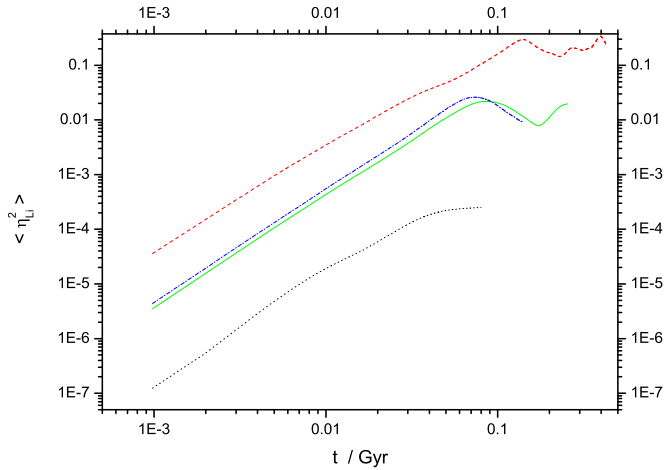,width=4in,height=3in,angle=0}}
\caption{\label{fig:bestfit}
$\sigma_{\eta_L}^2\equiv\langle\eta_{Li}^2\rangle$ as a function
of time, for stable spherical halos with different particle
numbers: $2\times10^4$ (dash/red line), $2\times10^5$(solid/green
line), $2\times10^6$(dot/black line) and the spherical evolving
halo with $2\times10^5$(dash dot/blue line) particles. }

\end{figure}

\section{Discussion and conclusion}
\label{sect:discussion}

{\it What is the difference?}--- Our demonstration with the two
numerical experiments brings out a physical difference in
microcosm between dark matter halos in simulations and in nature:
although they are both composed of collisionless CDM particles,
the process of gravitational encounter can be neglected in natural
halos, but not in simulation halos.

{\it How will it affect the simulation results?}--- On small
scale, the relaxation can greatly affect the behavior of
individual particles in simulation halos. Take the case of the
stable halo as an example. For the halo as a whole, the total
energy $E$ and the total angular momentum $L$ are each conserved.
What's more, statistically $\phi(r)$ is still a static and
spherically symmetric potential in a short period. However Fig 1
and Fig 2 clearly show changes of $E_i$ and $L_i$ caused by
gravitational encounters in a micro-statistical sense. In other
words, from the point of view of the individual particle of the
simulation halo, the potential is $no longer$ spherically
symmetric or static. To be exact, for a halo that comprises fewer
than approximately $10^6$ particles, gravitational collision must
be taken into count. The halo is no longer a system of
collisionless particles, and the collisionless Boltzmann Equation
is no longer applicable.

Will the general dynamical behavior of the whole system in the
large scale statistics respect, be affected by such microcosmic
effect? The answer is obviously yes. For the CDM halo in
simulation, the $E_i$ and $L_i$ of each particle can vary, which
means the distribution function $f(E_i,L_i)$ of the whole system
can be different from the natural case. And if we can't trust the
$f(E_i,L_i)$ in simulation, then the density profile $\rho(r)$ and
velocity dispersion of the halo we learn from the simulation will
be subject to doubt. Actually, whether we can express the
distribution function $f(\mathbf{x},\mathbf{v})$ in the form of
$f(E,L)$ needs to be considered (see the following discussion
about the Jeans Theorem). There isn't any mature theory dealing
with the effect of such microcosmic relaxation on $\rho(r)$ so
far, and more research is expected.

However, by comparing our results above with a previous research
on dynamic behavior of stellar clusters (Binney \& Tremaine 1987),
we can learn a lot about how the difference caused by the particle
number will affect the simulation results. Much research has been
done on stellar systems (such as globular clusters) which comprise
similar numbers of collisionless particles under pure
gravitational interaction. The dynamical effects caused by the
process of gravitational encounter in stellar systems can also
appear in simulation halos. A good example is the core collapse
process: due to star-star encounters in the core, some of the
stars get very high energies and evaporate away, while the core
shrinks as required by energy conservation. Following the way of
studying stellar systems, we can also estimate the time to core
collapse $\tau$ of one CDM halo (Binney \& Tremaine 1987, eq.
8-72) as:

\begin{equation}
\tau \simeq 20t_{rh} \simeq \frac{1.3\times10^{10}yr}{ln(0.4N)}
(\frac{M}{10^{12}M_{\odot}})^{1/2}(\frac{1M_{\odot}}{10^8
m})(\frac{r_h}{10kpc})^{3/2}
\end{equation}

Where $M$ and $r_h$ are the total mass and median radius of the
halo, $m$ is the mass of each particle of it, and $t_{rh}$ is the
median relaxation time of the system . Now we can see, the core
collapse time scale of an isolated spherical simulation CDM halo
with $M\simeq10^{12}M_{\odot}$,$r_h\simeq10kpc$ and including
$N\simeq10^4$ particles ($m=M/N\simeq10^8M_{\odot}$), will be
about the age of the universe. Based on the estimation above, one
might expect the CDM halo in simulation can avoid such core
collapse process. Unfortunately, considering:

(1) According to the hierarchical structure formation scenario in
a $\Lambda$CDM universe, small structures come into being first
and large halos comes from the merging of small ones.

(2) In the same simulation, with the same value of $m$ above but
much smaller $M$ and $r_h$, eq.(4) indicates small halos will have
a much shorter core collapse time scale than the age of the
universe(one can also see from Fig.4 and eq.(1)). So these small
halos in simulation will have to suffer the core collapse process,
and form one artificial cusp center. On the contrary, such core
collapse will not happen in a CDM halo in nature which includes
very large number of small $m\sim GeV$ particles, for they have a
time scale $\tau$ much longer than $10^{10}$ years.

(3)Numerical studies (Hayashi et al.2003, Michael \& Ma 2004) have
shown the cuspy centers can survive from merging process of both
the major mergers of galaxy halos and the evolution of
substructures.

Now we can figure out how do the bias affect the simulation
results: In a $\Lambda$CDM universe, small halos come into being
first and the unexpected core collapse process lead to the
artificial cuspy centers of them. Then these cuspy centers survive
from the merging process later and contribute to both the "cusp
problem" and the "substructure problem". No matter if there is
other explanation for these problems, the contribution of this
effects must be taken into account when we analyze simulation
results on small scales. One can even predict, with the
improvement of simulation technology, the "substructure problem"
can be even more serious if the particle number increased by only
one or two orders.

Last but not least, our discovery of the bias directly tests the
premise of the commonly used Jeans Theorem in galactic dynamics.
According to the Jeans Theorem, in a static and spherical
symmetric potential well, we can express the distribution function
$f(\mathbf{x},\mathbf{v})$ in phase space as $f(E,L)$ only when
the $E_i$ and $L_i$ of each particle are all integrals of motion.
Our result shows that, for the collisionless particle system with
pure gravitational interaction which has a particle number
$N\leqslant10^6$, gravitational encountering can not be dismissed.
As a result, the $E_i$ and $L_i$ for most of the particles will be
changed in a short period of time (see Fig.\,2). For such systems
like CDM halos in simulation and globular clusters, the $E_i$ and
$L_i$ are no longer integrals of motion and the Jeans Theorem is
no longer applicable, while it is still valid for CDM halos in
nature that are made up of a huge number of $GeV$ particles.

\begin{acknowledgments}

\begin{center}
\textbf{Acknowledgments}
\end{center}

We thank Professor T.Kiang for checking the manuscript, and we
thank Steen Hansen and Michael Boylan-Kolchin for helpful
discussion and suggestion.

\end{acknowledgments}

\clearpage

\end{document}